\documentclass[showpacs,twocolumn,showkeys,amsmath,amssymb,pra,superscriptaddress,nofootinbib]{revtex4-1}
 
\usepackage{graphicx} 

\begin{document}
\title{Nonlocal quantum superpositions of bright matter-wave solitons and dimers}

\author{Bettina Gertjerenken}
\email{b.gertjerenken@uni-oldenburg.de}
\affiliation{Institut f\"ur Physik, Carl von Ossietzky Universit\"at, D-26111 Oldenburg, Germany}

\author{Christoph Weiss}
\affiliation{Institut f\"ur Physik, Carl von Ossietzky Universit\"at, D-26111 Oldenburg, Germany}
\affiliation{Department of Physics, Durham University, Durham DH1 3LE, United Kingdom}

\date{24 August 2012}

\begin{abstract}
The scattering of bright quantum solitons at barrier potentials in one-dimensional geometries is investigated. Such protocols have been predicted to lead to the creation of nonlocal quantum superpositions. The centre-of-mass motion of these bright matter-wave solitons generated from attractive Bose-Einstein condensates can be analysed with the effective potential approach. An application to the case of two particles being scattered at a delta potential allows analytical calculations not possible for higher particle numbers as well as a comparison with numerical results. Both for the dimer and a soliton with particle numbers on the order of N = 100, we investigate the signatures of the coherent superposition states in an interferometric setup and argue that experimentally an interference pattern would be particularly well observable in the centre-of-mass density. Quantum superposition states of ultra-cold atoms are interesting as input states for matter-wave interferometry as they could improve signal-to-noise ratios.
\end{abstract}

\maketitle

\section{Introduction}
The experimental realisation of macroscopic quantum superpositions is a challenge of current research. Besides opening up a new realm for the study of fundamental quantum effects there would be possible applications in interferometry and quantum information science. As quantum superposition states are very sensitive to decoherence the creation of these fragile objects is a difficult task. So far experimental achievements include the realisation of mesoscopic superposition states on the single-atom level \cite{MunroeEtAl96}, a six-atom Schr\"odinger-cat state \cite{LeibfriedEtAl05} and superposition states in radiation fields \cite{BruneEtAl96}.  Further systems of interest for the creation of many-particle entanglement are cavity quantum optomechanical systems \cite{RomeroIsartEtAl11} and mesoscopic superpositions of internal degrees of freedom. In the field of ultra-cold atoms mesoscopic superposition states are important for matter-wave interferometry as they allow improved signal-to-noise ratios \cite{GiovannettiEtAl2004}. Many-particle entanglement in macroscopic systems enables investigations of the quantum-classical boundary and decoherence mechanisms \cite{HarocheRaimond06}. Therefore larger interferometric objects are required. While interferences have been observed with fullerenes \cite{HornbergerEtAl03,ArndtEtAl99} and large organic molecules \cite{GerlichEtAl2011} the creation of a macroscopic nonlocal Schr\"odinger-cat state of ultra-cold atoms still is a challenge. Recent proposals have been the creation of a nonlocal state of a Bose-Einstein condensate suspended between the two wells of a two-site optical lattice \cite{CarrEtAl10} or via scattering of bright matter-wave solitons (BMWS) \cite{WeissCastin09,StreltsovEtAl09}. Single BMWS \cite{KhaykovichEtAl02} and soliton trains \cite{StreckerEtAl02,CornishEtAl06} have already been realised experimentally in low-dimensional attractive Bose gases. Theoretical investigations of one-dimensional attractive Bose-gases \cite{CarrBrand04,CastinHerzog01} comprise the investigation of BMWS collisions \cite{BillamEtAl11} and collision-induced entanglement \cite{LewensteinMalomed09}.

In one-dimensional geometries the scattering of a BMWS or a tightly bound dimer off a barrier potential like a laser focus has been predicted to lead to the creation of a nonlocal quantum superposition of all the atoms being located to the left and to the right of the scattering potential. In a Fock space representation with states $|n_{\mathrm{L}},n_{\mathrm{R}}\rangle$, where $n_{\mathrm{L}}$ ($n_{\mathrm{R}}$) denotes the particles left (right) of the barrier, this would correspond to the NOON-state \cite{Wilfeuer07}
\begin{equation}\label{eq:NOON}
|\Psi_{\mathrm{NOON}}\rangle = \frac{1}{\sqrt{2}}\left(|N,0 \rangle + \mathrm{e}^{\mathrm{i}\phi}|0,N \rangle\right)
\end{equation}
where $\phi$ is the relative phase between the two parts of the superposition. Experimental parameters for this proposed realization of the ideal of a nonlocal quantum superposition have been estimated in \cite{WeissCastin09}: for a BMWS consisting of $N\cong 100$ $^7$Li atoms with a typical size of about 1 $\mu$m a spatial separation of about 100 $\mu$m is found to be reachable. The initial wave-function should be prepared carefully such that the beam-splitter acts on the soliton as a whole and not on each single atom. Using the effective potential approach from \cite{WeissCastin09} it is possible to investigate the dynamics of the centre of mass (CoM). While in Ref [11] the effective potential approach has been applied to broader potentials, here it will be applied for smaller potentials which is more interesting from the point of view of a theoretical physicist: for low particle numbers, it allows to perform Rayleigh-Schr\"odinger perturbation theory beyond the leading order of the effective potential investigated in Ref.~\cite{WeissCastin09}. This thus renders a detailed test of the effective-potential approach possible.

Two-particle bound states are interesting both experimentally \cite{WinklerEtAl06} and theoretically \cite{PiilMolmer07,PetrosyanEtAl07,JavanainenEtAl10,SoerensenEtAl12}. An application of the effective potential approach \cite{WeissCastin09} to the case of a tightly-bound dimer being scattered off a delta potential is particularly interesting: contrary to higher particle numbers $N \ge 3$ analytical calculations of the effective potential, even in higher orders, still are possible with reasonable effort and can be compared with numerical results. These can be obtained via discretisation of the Schr\"odinger equation. The scattering of bright solitons of delta function like potentials has also been investigated on the mean-field level and has been found to successfully describe many aspects of the dynamics \cite{KivsharEtAl89,KivsharEtAl90,AkkermansEtAl08}. The nonlinear Gross-Pitaevskii equation is not investigated here as it does not allow superposition states which are in the focus of this work.

Considering the nonlocal quantum superposition (\ref{eq:NOON}) in an experiment, one should always find all particles on the same side of the scattering potential. As by itself this is not yet a convincing signature of a quantum superposition rather than a statistical mixture, we will investigate the subsequent recombination and interference of the two coherent parts of the quantum superposition. The combined measurement of both signatures would be a demonstration for the existence of a Schr\"odinger-cat state. Interference experiments with matter waves can be of two very different kinds \cite{HarocheRaimond06}: while the first one involves the interference of two independent condensates \cite{AndrewsEtAl97}, the other one is based on the splitting of a single system into two components. In the interference of two independent condensates \cite{AndrewsEtAl97} an interference pattern is visible in a single run of the experiment but will eventually wash out by averaging over many realisations due to the random phase between the condensates. Here we investigate a complementary kind of interference where the interference pattern will \emph{build up} in a sequence of runs: The density distribution for the CoM $X=\frac{1}{N}\sum_{j = 1}^N x_{j}$ of the system is given by
\begin{equation}\label{eq:CoMdensity}
\rho_{\rm{CoM}} (x)= \langle \delta\left(x-X\right) \rangle,
\end{equation}
where $x$ denotes the spatial coordinate, whereas the single-particle density reads
\begin{equation}\label{eq:onedensity}
\rho_{\rm{one}}(x) = \frac{1}{N}\sum_{j}\langle \delta\left(x-x_j\right) \rangle
\end{equation}
with the position $x_j$ of the particles. While there are set-ups for which a single experiment yields an interference pattern~\cite{AndrewsEtAl97}, in general quantum mechanical expectation values like equations (\ref{eq:CoMdensity}) and (\ref{eq:onedensity}) only make statements about an average of many runs of the experiment. For the CoM density (\ref{eq:CoMdensity}), a single experiment will only give a point (the CoM of all atoms measured). As for the double-slit experiment with single photons \cite{GravierEtAl86}, the interference pattern only builds up after repeating the experiment many times. The same experiments can also be used to determine the single-particle density (3): here the positions of all atoms from all experiments are added. Here we assume adequate experimental stability to garantuee that the shift of the interference pattern from run to run is smaller than the distance of neighboring interference maxima to avoid washing out of the pattern:  an experiment will have to combine the low temperatures of Ref.~\cite{LeanhardtEtAl03} (to prepare the initial state, cf.~\cite{WeissCastin09}) with the vacuum of Ref.~\cite{AndersonEtAl95} (to suppress decoherence introduced by free atoms) and the particle-number control available in Experiments like~\cite{GrossEtAl10}.

The paper is organized as follows: In the next section we will introduce the Lieb-Liniger model \cite{LiebLiniger63,McGuire64} for the description of a one-dimensional attractive Bose gas and review the effective potential approach from \cite{WeissCastin09}. The third section covers numerical investigations of the scattering of a tightly bound dimer at a delta potential and the resulting interference patterns as well as the application of the effective potential to this two-particle case. In the fourth section the considerations about interference patterns in the CoM and single-particle density are extended to the experimentally relevant $N$-particle case.

\section{Models}
\subsection{Lieb-Liniger model of a one-dimensional attractive Bose gas}\label{sec:model}
One-dimensional Bose gases can be described in the exactly solvable Lieb-Liniger model \cite{LiebLiniger63,McGuire64} where the many-body Hamiltonian for $N$ bosons of mass $m$ interacting via contact interaction with coupling constant $g_{1\rm{d}}$ is given by
\begin{equation}\label{eq:H0}
\hat{H_{0}} = \sum_{j=1}^N \frac{p_{j}^2}{2m} + g_{1\rm{d}}\sum_{j<\ell}\delta(x_{j} - x_{\ell}).
\end{equation}
The resulting Schr\"odinger equation is separable in CoM and relative coordinates and can be diagonalised with the Bethe ansatz. For attractive interaction with $g_{1\rm{d}}<0$ bound states with negative energies are expected. The ground state of the system is the quantum soliton
\begin{equation}\label{eq:groundstate}
\psi_{0}(\mathbf{x}) = C_N\exp\left(-\beta \sum_{1\leq j < \ell \leq N} |x_{j} - x_{\ell}| \right),
\end{equation}
where $\beta = -mg_{1\rm{d}}/2\hbar^2 > 0$ and $C_N =\left[\frac{(N-1)!}{N}(2\beta)^{N-1}\right]^{1/2}$ \cite{CastinHerzog01}. For $N\gg1$ the single-particle density (\ref{eq:onedensity}) for the CoM being localized around $x=0$ is well approximated by the mean-field density profile \cite{CalogeroDegasperis75}
\begin{eqnarray} \label{eq:rhosoliton}
\rho_N\left(x\right) & =  \int \mathrm{d}x_1\ldots\mathrm{d}x_N |\psi_{0}(\mathbf{x})|^2\delta \left(\frac{x_1 + \dots + x_N}{N} \right)\delta \left(x_1-x \right) \nonumber \\
& =  \frac{mg}{4\hbar^2}N^2\left[\cosh\left( \frac{mg_{1\rm{d}}}{2\hbar^2}Nx\right) \right]^{-2}.
\end{eqnarray}
Thus, $x$ represents the distance from the CoM of the soliton. Note that the wave-function (\ref{eq:groundstate}) already is symmetric with respect to any permutation of particles. The ground state (\ref{eq:groundstate}) of the Hamiltonian (\ref{eq:H0}) has energy
\begin{equation}\label{eq:energyofsoliton}
E_0\left( N \right) =-\frac{1}{24}\frac{mg^2_{1\rm{d}}}{\hbar^2}N\left(N^2-1\right)
\end{equation}
and is separated by an energy gap
\begin{eqnarray}\label{eq:gap}
|\mu| &= E_0\left(N-1\right) - E_0\left(N\right) \nonumber \\ &= \frac{mg_{1\rm{d}}^2N\left(N-1\right)}{8\hbar^2}
\end{eqnarray}
from a continuum of solitonic fragments. The corresponding $N$-particle solutions with CoM wave vector $K=Nk$, total mass $M=Nm$ and 
energies
\begin{equation}\label{eq:energysol}
E_K(N) = E_0(N) + \frac{\hbar^2K^2}{2M}.
\end{equation}
read
\begin{equation}\label{eq:solitonwavefunction}
\psi_{N,k}(\mathbf{x}) = \psi_{0}(\mathbf{x}) \exp\left(\mathrm{i} k\sum_{j = 1}^N x_{j}\right).
\end{equation}

Hence, for low CoM kinetic energies $E_K(N)-E_0(N)<|\mu|$ the scattering of the BMWS at an additional scattering potential 
\begin{equation}\label{eq:scatpot}
V(\mathbf{x}) = \sum_{j=1}^N \tilde{V}\left( x_{j} \right),
\end{equation}
leading to the total Hamiltonian $H = H_0 + V(\mathbf{x})$, will be elastic. In this work we assume the scattering potentials to be delta distributions $\tilde{V}\left( x_{j} \right) \propto \delta\left(x_{j}\right)$ which will be motivated in section \ref{sec:ScatteringDelta}. As the soliton cannot break into parts nonzero probability amplitudes of the scattered wave-function on both sides of the scattering potential imply the generation of a Schr\"{o}dinger-cat state. For higher energies particle excitations are possible and the soliton can break up into several 'lumps' of bound particles. Those excited states can be found in a generalised Bethe ansatz \cite{CastinHerzog01}. For $N=2$ they read
\begin{equation}\label{eq:excstatespm}
|\Psi\rangle_{\pm} \propto \left[\cos\left(kx_{\rm{r}} \right)-\frac{\beta}{k} \sin\left(k|x_{\rm{r}}|\right) \right]\mathrm{e}^{\pm\mathrm{i}KX}
\end{equation}
with CoM and relative coordinates
\begin{eqnarray}\label{eq:CoMandrelcoord}
X&=\frac{x_1+x_2}{2},\, &K=k_1+k_2,\, \nonumber \\ x_{\rm{r}} &= x_2-x_1,\, &k=\frac{k_1-k_2}{2}.
\end{eqnarray}

\subsection{Effective potential approach}\label{sec:effpotapproach}
Adding the additional scattering potential (\ref{eq:scatpot}) to the Hamiltonian (\ref{eq:H0}) renders the Schr\"{o}dinger equation nonintegrabel via the Bethe ansatz. Nonetheless, for suitably smooth scattering potentials and low kinetic energies the CoM motion can again be solved independently of the relative motion in the mathematically justified effective potential approach \cite{WeissCastin09}. It is essentially based on the existence of the energy gap (\ref{eq:gap}). To solve the CoM simply replacing the exact potential (\ref{eq:scatpot}) by $NV(X)$ is not sufficient, see section \ref{sec:effectivepot}. Instead the Schr\"odinger equation for the CoM motion is given by
\begin{equation}\label{eq:CoMSGL}
H_{\mathrm{eff}} = -\frac{\hbar^2}{2M}\partial_X^2 + V_{\mathrm{eff}}\left( X\right)
\end{equation}
where $X=\frac{1}{N}\sum_{j = 1}^N x_{j}$ is the CoM of the particles and the effective potential is the convolution of the internal density profile of the ground state soliton with the barrier potential:
\begin{equation}\label{eq:Veff}
V_{\mathrm{eff}}(X) = \int \mathrm{d}^Nx\delta\left(X - \sum_{j = 1}^N x_{j} \right) |\psi_{N,k}(\mathbf{x})|^2 \sum_{j =1}^N  \tilde{V}\left( x_{j} \right).
\end{equation}

\section{Scattering of a dimer off a delta potential}\label{sec:ScatteringDelta}
In an experiment single atoms could serve as scattering potentials, in the following mathematically simplified as delta distributions. We therefore investigate the scattering of a dimer off such a delta potential in a one-dimensional geometry and discuss signatures of the resulting Schr\"odinger-cat states. Moreover the effective potential approach of section \ref{sec:effpotapproach} will be applied to this situation which is mathematically more challenging than the case of broader potentials. Contrary to higher particle numbers, for two particles analytical calculations nonetheless still are possible and can furthermore be compared with numerical results obtained via discretising the Schr\"odinger equation.

Within a Bose-Hubbard Hamiltonian \cite{GreinerEtAl02b,JakschZoller05}
\begin{eqnarray}
\hat{H}_{\rm{lattice}} = &-J \sum_{j}\left(\hat{a}_{j}^{\dagger}\hat{a}_{j+1}^{\phantom{\dagger}} + \hat{a}_{j+1}^{\dagger}\hat{a}_{j}^{\phantom{\dagger}}  \right) \nonumber \\ &+ \frac{U}{2}\sum_{j} \hat{n}_{j}\left(\hat{n}_{j}-1  \right), \ \ \ U<0,
\end{eqnarray}
describing the dynamics of $N$ bosonic particles on a lattice, it is possible to use exact eigen-states. Here, $J$ is the tunnelling strength, $U$ denotes the pair interaction, $\hat{a}_{j}^{\dagger}$ and $\hat{a}_{j}^{\phantom{\dagger}}$ are the usual boson creation and annihilation operators and $\hat{n}_{j}=\hat{a}_{j}^{\dagger}\hat{a}_{j}^{\phantom{\dagger}}$ is the number operator for site $j$. For a single particle on a lattice the dispersion relation is given by 
\begin{equation}\label{eq:disprelone}
E_{\rm{one}} = -2J\cos(kb).
\end{equation}
Comparing equation (\ref{eq:disprelone}) with the dispersion relation for suitably small lattice step sizes $b$ under consideration of the lower band edge $-2J$ and choosing $Jb^2 = \frac{\hbar^2}{2m}$ then renders a numerical investigation of the free particle motion possible. In the case of a dimer the two-particle bound states \cite{ValienteEtAl08,WeissBreuer09,WinklerEtAl06,PiilMolmer07,PetrosyanEtAl07,JavanainenEtAl10} are given by
\begin{equation}
|\Psi\rangle = \sum_{j,\ell} c_{j \ell} |j \rangle |\ell \rangle 
\end{equation}
with
\begin{equation}\label{eq:wavefunctions}
c_{j \ell} =
\begin{cases}
x_{-}^{|j-\ell|} \exp\left[\mathrm{i}kb\left(j + \ell \right) \right]   &, j \neq \ell \\
\frac{1}{\sqrt{2}}\exp\left[\mathrm{i}kb\left(j + \ell\right) \right]   & , j = \ell 
\end{cases}
\end{equation}
and
\begin{equation}
x_{-} = \sqrt{\frac{U^2}{16J^2\cos^2(kb)} + 1} - \frac{|U|}{|4J|\cos(kb)}.
\end{equation}
This leads to a dispersion relation
\begin{equation}
E_{\rm{two}} \simeq-4J\left( \frac{U^2}{16J^2} +1 \right)^{1/2} + \frac{8J^2b^2}{\left(16J^2 + U^2\right)^{1/2}}k^2.
\end{equation}
Analogue to the one-particle case for small values of $b$ and under consideration of the offset $-4J$ a comparison with the energy 
\begin{equation}
E_{N=2,\rm{soliton}} = -\frac{\hbar^2\beta^2}{m}+\frac{\hbar^2k^2}{m}
\end{equation}
of a two-particle soliton according to equation (\ref{eq:energysol}) then enables the numerics for a tightly bound dimer moving in a one-dimensional geometry using the dimer wave-functions in a tight-binding lattice.

In the following, like in figure~\ref{fig:comdensity}~(a), the initial wave function with mean momentum $k_0$ and width $a$ is built up of the two-particle eigenfunctions (\ref{eq:wavefunctions}) where $k$ should be restricted to $0 < k <\beta$ so that the dimer cannot break apart for energetic reasons. We choose the upper integral boundary $k_{\rm{max}}$ to be about twice the mean momentum:
\begin{eqnarray}\label{eq:initialwave}
\Psi_0\left(x_1,x_2;t=0\right) &\propto \int_0^{k_{\rm{max}}} \mathrm{d}k \exp\left(-\frac{a^2\left(k-k_0 \right)^2}{2} \right) \nonumber \\ & \;\;\;\;\;\;\;\;\; \times\Psi_{2,k}\left(x_1-x_0,x_2-x_0\right).
\end{eqnarray}

\subsection{Interference patterns}\label{sec:numerics}
In the low energy-regime the scattering of a dimer at a delta potential
\begin{equation}\label{eq:exactpotential}
V(x_1,x_2) = v_0\left[\delta(x_1) + \delta(x_2)\right],
\end{equation}
located in the middle of the lattice, can lead to the creation of nonlocal mesoscopic superposition states where both particles always should be found on the same side of the scattering potential. To probe the existence of such a Schr\"odinger-cat state we thus use the criterion \cite{Weiss10}
\begin{equation}\label{eq:miaou}
p_{\mathrm{miaou}} = \int_{-\infty}^0 \mathrm{d}x_1 \int_{0}^{\infty} \mathrm{d}x_2 \left|\Psi\left(x_1,x_2;t \right)  \right|^2
\end{equation}
and ensure that it takes on sufficiently low values on the order of $10^{-8}$. To investigate the interference patterns the contrast
\begin{equation}\label{eq:contrast}
C = \frac{I_{\rm{max}}-I_{\rm{min}}}{I_{\rm{max}}+I_{\rm{min}}}
\end{equation}
will be calculated on a suitable interval in the middle of the pattern.

\begin{figure}
\includegraphics[width=1.0\columnwidth]{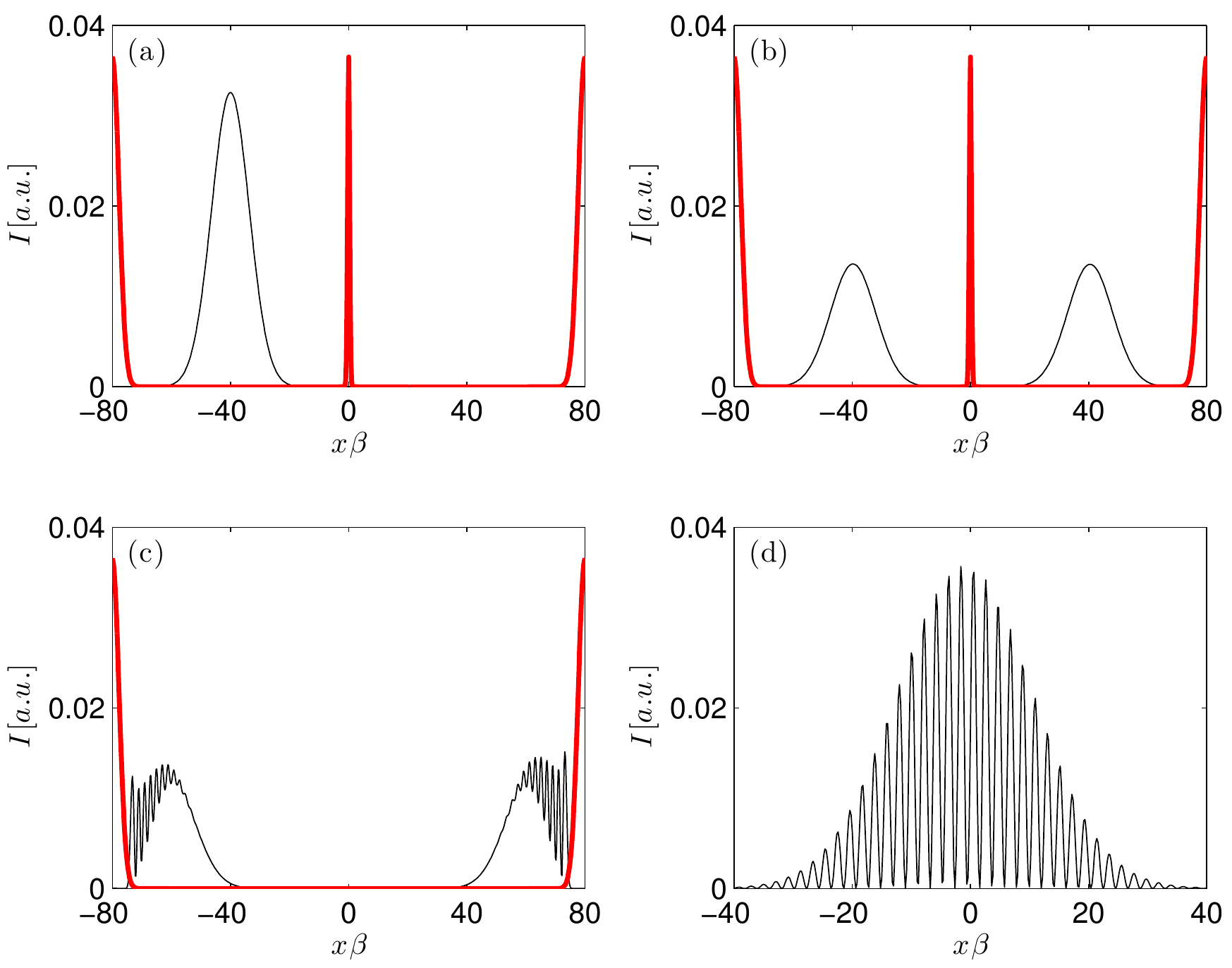}
\caption{\label{fig:comdensity} CoM density (thin line) for the scattering of a dimer at a delta potential $V(x_1,x_2) = v_0\left[\delta(x_1) + \delta(x_2)\right]$. The delta potential situated in the middle of the lattice as well as the reflection potentials at the edges are displayed schematically (thick lines). Lattice size: 801x801 lattice points. (a) Initial wave function with width $a\beta = 18$ and mean wave vector $k_0/\beta = 0.75$ centred around $x_0\beta=-40$. (b) Schr\"odinger-cat state with $p_{\mathrm{miaou}} = 0.25\cdot10^{-8}$ and 50\%-50\% splitting, shown at $t\hbar\beta^2/m = 136$. (c)~Reflection of the two parts of the quantum superposition at reflection potentials on the edges. Scattering potential is switched off. (d) Recombination of the two parts of the wave function leading to an interference pattern in the CoM coordinate with maximal contrast $C = 1$ at $t\hbar\beta^2/m = 255$. The contrast is calculated using equation (\ref{eq:contrast}) on the interval $x\beta \in [-10,10]$.}
\end{figure}

Figure \ref{fig:comdensity} shows numerical results for the CoM density distribution (\ref{eq:CoMdensity}): The initial wave function in figure~\ref{fig:comdensity} (a) is implemented according to (\ref{eq:initialwave}). Scattering at the delta potential then leads to the creation of a nonlocal superposition, displayed in figure~\ref{fig:comdensity} (b). To allow for a subsequent recombination of the wave packets reflection potentials are added at the edges of the lattice. After being reflected, cf. figure~\ref{fig:comdensity} (c), an interference pattern with maximal contrast $C=1$ arises in the CoM density which can be seen in figure~\ref{fig:comdensity} (d). The contrast has been calculated on the interval $x\beta \in [-10,10]$.

\begin{figure}
\includegraphics[width=1.0\columnwidth]{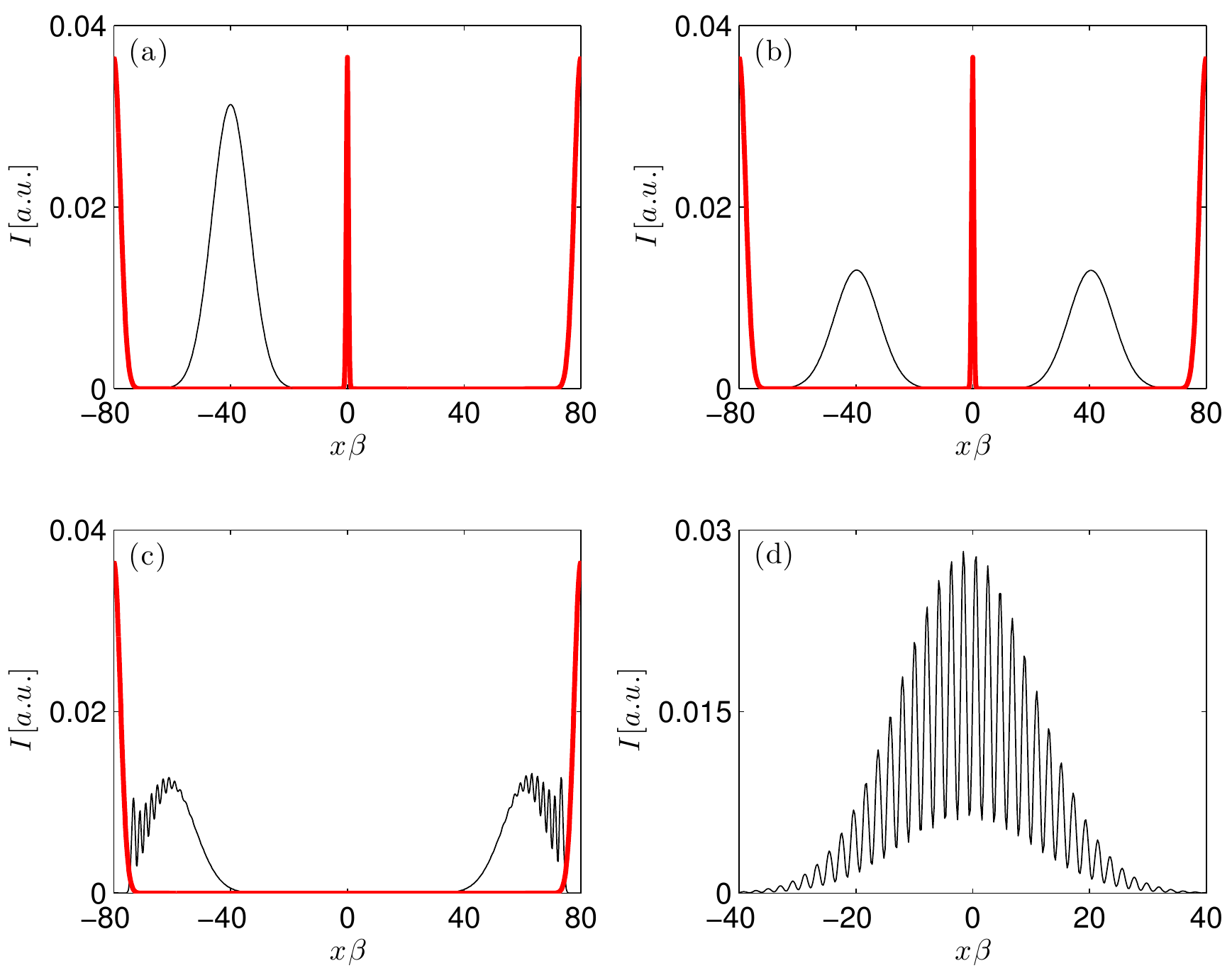}
\caption{\label{fig:onedensity} Single-particle density (thin line) for the scattering of a dimer at a delta potential (thick line). Same parameters  and set-up as in figure \ref{fig:comdensity}. (a) Initial wave function. (b) Schr\"odinger-cat state with $p_{\mathrm{miaou}} = 0.25\cdot10^{-8}$ and 50\%-50\% splitting, shown at at $t\hbar\beta^2/m = 136$. (c) Reflection of the two parts of the quantum superposition at reflection potentials on the edges (thick line). Scattering potential is switched off. (d) Recombination of the two parts of the wave function leading to an interference pattern with reduced contrast $C = 0.750$ in the single-particle density, shown at $t\hbar\beta^2/m = 255$ and computed as in figure~\ref{fig:comdensity}.}
\end{figure}

The results for the single-particle density (\ref{eq:onedensity}) which are displayed in figure~\ref{fig:onedensity} differ clearly from this. The contrast of the interference pattern is reduced to about 0.75, again calculated for $x\beta \in [-10,10]$.

Summarising, the scattering of a dimer off a delta potential can lead to the creation of a Schr\"odinger-cat state. While the subsequent recombination of the two coherent parts of the wave-function leads to an interference pattern with maximum contrast in the CoM density the visibility of the pattern in the single-particle density is reduced. These numerical results will in section \ref{sec:interference} be compared with analytical considerations which can be extended to the $N$-particle case.\\

\subsection{Effective potential approach for the scattering of a dimer at a delta potential}\label{sec:effectivepot}

The existence of the energy gap (\ref{eq:gap}) allowed the creation of the mesoscopic superposition state in \ref {sec:numerics} and renders the effective potential approach from section \ref{sec:effpotapproach} possible. For the scattering of a dimer off a delta potential it is possible to compare the effective potential approach even in higher order perturbation theory with numerical results using again the tight-binding wavefunctions of section \ref{sec:numerics}.

The exact potential (\ref{eq:exactpotential}) can be expressed as
\begin{equation}\label{eq:potperturbed}
V(x_1,x_2) = V_X(X) + \Delta V(x_1,x_2)
\end{equation}
where $V_X(X) = 2v_0\delta\left( X \right)$ and
\begin{equation}
\Delta V(x_1,x_2) = v_0\left[- 2\delta\left( X \right) + \delta\left(x_1\right)  + \delta\left(x_2\right) \right].
\end{equation}
This perturbation $\Delta V(x_1,x_2)$ to the CoM contribution $V_X(X)$ induces an energy shift $\Delta E$ on the CoM motion. The Schr\"odinger equation for the CoM wave-function $\Psi(X)$ then is given by
\begin{equation}
\mathrm{i}\hbar \partial_t |\Psi(X)\rangle= \left[-\frac{\hbar^2}{2M}\partial_X^2 + V_X(X) + \Delta E(X)\right]|\Psi(X)\rangle.
\end{equation}
The energy shift $\Delta E$ on the CoM motion due to the relative motion can be computed in Rayleigh-Schr\"odinger perturbation theory. In first order it is given by
\begin{equation}\label{eq:firstorder}
\Delta E^{(1)}(X) = \langle \Psi_{0,N=2}^{(0)}|\Delta V(x_1,x_2)| \Psi_{0,N=2}^{(0)} \rangle.
\end{equation}
with the ground state soliton (\ref{eq:groundstate}) for two particles,
\begin{equation}\label{eq:densityN2}
\Psi_{0,N=2}^{(0)}(x_{\rm{r}}) =  \sqrt{2\beta}\exp(-\beta |x_{\rm{r}}|),
\end{equation}
using CoM and relative coordinates (\ref{eq:CoMandrelcoord}). As shown in \ref{app:effpot}, the energy correction adopts the form
\begin{equation}
\Delta E^{(1)}(X) = -2v_0 \delta(X) + 4\beta v_0 \exp(-4\beta |X|)
\end{equation}
such that up to first order perturbation theory the effective potential in the Schr\"odinger equation (\ref{eq:CoMSGL}) for the CoM can be identified as the convolution
\begin{eqnarray}\label{eq:effpot1}
V_{\mathrm{eff}}^{(0)}(X) + V_{\mathrm{eff}}^{(1)}(X) &=V_X(X) +\Delta E^{(1)}(X) \nonumber \\ &= 4\beta v_0 \exp(-4\beta |X|)
\end{eqnarray}
of the scattering potential (\ref{eq:exactpotential}) with the density profile (\ref{eq:densityN2}) of the dimer. Taking into account excited states \cite{CastinHerzog01} gives the higher order corrections
\begin{equation}\label{eq:Veff2}
V_{\mathrm{eff}}^{(2)}(X)=-\frac{2v_0^2m}{\hbar^2}\exp(-4\beta|X|)\left[1-8\beta|X|\exp(-4\beta|X|) \right]
\end{equation}
and
\begin{eqnarray}\label{eq:Veff3}
V_{\mathrm{eff}}^{(3)}(X) =&16v_0^3 \frac{m^2}{\hbar^4}\exp(-4\beta|X|)\nonumber \\&\times\left[ C_1(X) + C_2(X)\exp(-4\beta|X|)\right. \nonumber \\ &\left.+C_3(X)\exp(-8\beta|X|) \right]
\end{eqnarray}
to the effective potential (cf. \ref{app:effpot}), where $C_1(X)=\frac{1}{16\beta}+ \frac{1}{8\beta^2}\delta(X)$, $C_2(X)=- \left(|X|+\frac{1}{4\beta} +\frac{1}{16\beta^2}\delta\left(X \right)\right)$ and $C_3(X)=\left(8|X|^2\beta +|X| +\frac{1}{8\beta}\right)$.

\begin{figure}[h]
\begin{center}
\includegraphics[width = 1.0\linewidth]{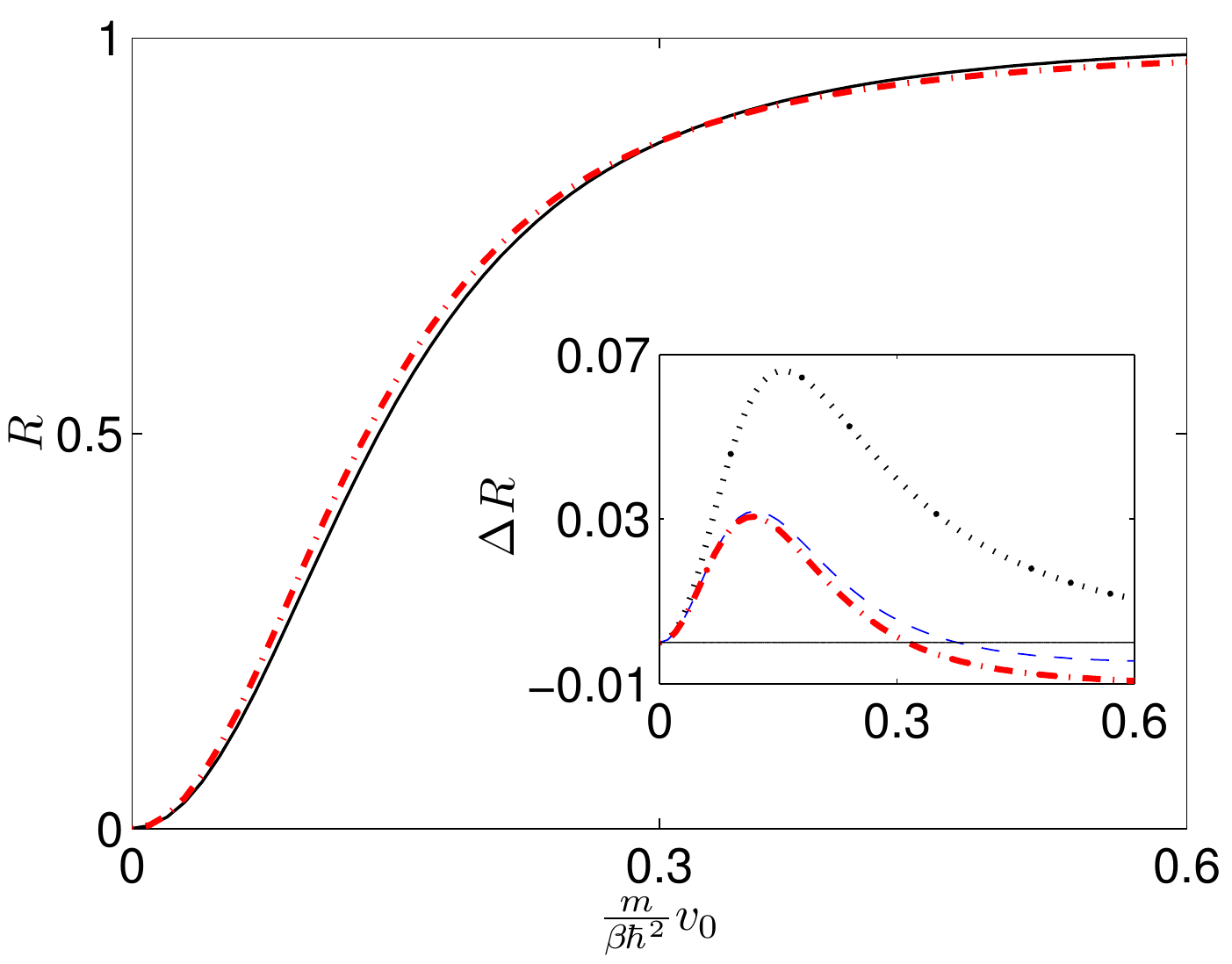}
\end{center}
\caption{\label{fig:rversusv0np801} Reflection coefficient $R(v_0)$ for the scattering of a dimer at a delta potential $V(x_1,x_2) = v_0\left[\delta(x_1) + \delta(x_2)\right]$, calculated on a lattice with 801x801 lattice points and $-160 < x\beta < 160$. The initial wave function with width $a\beta = 28$ and mean momentum $k_0/\beta=0.375$ was located at $x_0\beta = -80$. Numerics for exact potential (solid line) and effective potential up to 3rd order perturbation theory (dash-dotted line). Taking only the zeroth order corrections into account leads to full transmission over the whole range of parameters and hence does not describe the dynamics correctly at all. Inset: Deviation $\Delta R=R_{\delta}(v_0)-R_{\rm{eff}}(v_0)$ including up to 1st order corrections (dotted line), 2nd order corrections (dashed line) and 3rd order corrections (dash-dotted line).}
\end{figure}

In the following the scattering of a dimer at a delta potential will be investigated numerically for suitable parameters, on the one hand using the exact potential (\ref{eq:exactpotential}) and on the other hand the results for effective potential. The scattering potential is again located in the middle of a lattice with spatial size $L$ where $-L/2 < X <  L/2$ and the initial wave-function is chosen according to (\ref{eq:initialwave}) to ensure the dimer not to break apart. Figure \ref{fig:rversusv0np801} shows the reflection coefficient
\begin{equation}
R(v_0)= \int_{-L/2}^0 \mathrm{d}X |\Psi(X)|^2,
\end{equation}
which is increasing with the strength of the scattering potential $v_0$. Neglecting the correction $\Delta E(X)$ leads to full transmission over the whole range of values for the height of the scattering potential and thus in zeroth order the dynamics is not described correctly at all. As can be seen in figure~\ref{fig:rversusv0np801} the reflection coefficient $R_{\delta}(v_0)$ for the exact potential is well-approximated in third order perturbation theory. In the inset it is shown that already in first order the dynamics are qualitatively well-described. Especially in the experimentally interesting regime with $R\approx0.5$ a convergence towards $R_{\delta}(v_0)$ can be observed: the second order leads to a clear improvement while the additional contribution in third order is small. For larger potential heights with $R>0.8$ taking into account only up to 2nd rather than 3rd order contributions leads to better results.

\section{Interference patterns as signatures of quantum superposition states}\label{sec:interference}
In section \ref{sec:numerics} the interference of the two coherent parts of the quantum superposition has been investigated. In the following we will analytically explain the different values of the contrast in the CoM and the single-particle density and extend the two-particle results to considerations for $N$ particles. We will omit the envelopes of the wavepackets and restrict to the interference of two plane waves with opposed CoM wave vector $K$. If already in this case no interference is observed neither with wave-packets built up of plane waves this would be the case.

For the description of a dimer we use again the relative and CoM coordinates (\ref{eq:CoMandrelcoord}). The wave function (\ref{eq:solitonwavefunction}) then reads
\begin{equation}
\Psi_{\pm} = \sqrt{2\beta}\mathrm{e}^{-\beta |x_{\rm{r}}|}\mathrm{e}^{\pm\mathrm{i}KX}.
\end{equation}
The interference of the two parts $\Psi_+$ and $\Psi_-$ of the superposition
\begin{equation}
\Psi_+ + \Psi_- =2\sqrt{2\beta}\mathrm{e}^{-\beta |x_{\rm{r}}|}\cos\left(KX\right)
\end{equation}
results in the following interference pattern in the CoM-density distribution (\ref{eq:CoMdensity}) 
\begin{eqnarray}
\langle \delta\left(x-X \right) \rangle_{\rm{gr}} & =  \int_0^{\infty}\mathrm{d}x_{\rm{r}}\int_{-\infty}^{+\infty}\mathrm{d}X|\Psi_+ + \Psi_-|^2\delta\left(x-X \right) \nonumber \\ & \propto \cos^2\left(xK \right).
\end{eqnarray}
Hence, in the CoM-coordinate an interference pattern with fringe spacing $\pi/K$ can always be observed and the contrast (\ref{eq:contrast}) takes on its maximal value $C_{\rm{gr,CoM}}=1$.

\begin{figure}
\includegraphics[width=1.0\columnwidth]{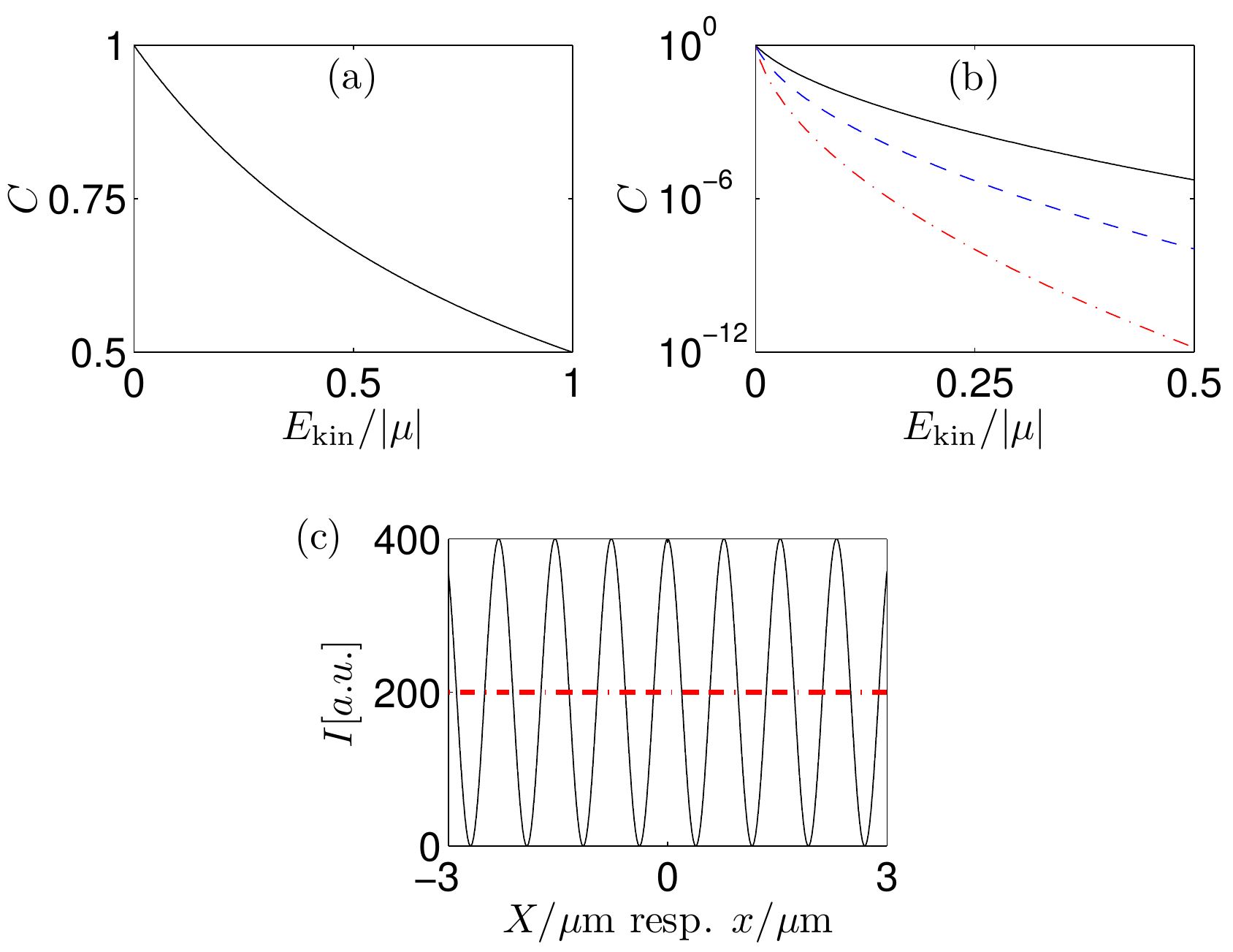}
\caption{\label{fig:contrastN} Contrast~(\ref{eq:contrast}) for the interference pattern in the single-particle density of the coherent parts of a Schr\"odinger-cat state after the scattering of a tightly bound dimer respectively a bright soliton at a delta potential. (a) Contrast~(\ref{eq:contrastonegr}) depending on the ratio of kinetic and binding energy for the dimer. (b) Contrast~(\ref{eq:contrastNoneparticle}) for a soliton with $N=50$ (solid line), $N=100$ (dashed line) and $N=200$ (dash-dotted line) particles. (c) CoM interference (solid line) of two counterpropagating plane waves $\Psi(X)\sim \exp(\mathrm{i}KX)$ and $\Psi(X)\sim \exp(-\mathrm{i}KX)$ for $N=100$ $^7$Li atoms with \mbox{$\hbar K/M \simeq 0.37$ mm/s.} In the corresponding single particle density (dash-dotted thick line) no interference pattern can be observed for these typical experimental parameters.}
\end{figure}

For the single-particle density (\ref{eq:onedensity}) the corresponding expectation value is
\begin{eqnarray}\label{eq:singlepartN2}
\rho_{\rm{gr,one}}& = \frac{1}{2}\sum_{j=1,2} \langle \delta \left( x - x_j  \right) \rangle \nonumber \\ & \propto \frac{1}{\left(4\beta^2 +K^2\right)}\left[4\beta^2\cos\left(2Kx \right) +4\beta^2 +K^2\right]
\end{eqnarray}
and as the ratio $\beta/|K|$ is related to the ratio of binding to CoM kinetic energy via $2\beta/|K|=\sqrt{\frac{|\mu|}{E_{\rm{kin}}}}$ the contrast now is given by
\begin{equation}\label{eq:contrastonegr}
C_{\rm{gr,one}} =\frac{1}{1+{\frac{E_{\rm{kin}}}{|\mu|}}}.
\end{equation}
For low CoM kinetic energies the contrast is maximal while for high values of $E_{\rm{kin}}$ the interference vanishes:
\begin{eqnarray}
\lim_{\frac{E_{\rm{kin}}}{|\mu|} \to 0}C_{\rm{gr,one}} &= 1, \\
\lim_{\frac{E_{\rm{kin}}}{|\mu|} \to \infty}C_{\rm{gr,one}} &= 0.
\end{eqnarray}
This agrees qualitatively with our numerical observations, cf. figure \ref{fig:onedensity}. The dependence of the contrast on $E_{\mathrm{kin}}/|\mu|$ is shown in figure~\ref{fig:contrastN} (a).

For CoM kinetic energies larger than the binding energy in an experiment excitations could occur. What would be the influence on the visibility of the interference pattern in the CoM density? Repeating the same considerations as for the ground state for the excited states (\ref{eq:excstatespm}) leads again to an interference pattern 
\begin{equation}
\lim_{L \to \infty}\langle \delta\left(x-X \right) \rangle_{\rm{exc}} \propto \cos^2\left(Kx\right)
\end{equation}
with maximal contrast $C_{\rm{exc,CoM}} = 1$, cf. \ref{eq:AppA}. For the ground state we discovered the same result and looking separately at ground or excited wave functions additional phase factors are irrelevant for the contrast. However, in an experiment involving excitations phase factors could become important. Hence a superposition of the patterns $\cos^2\left(Kz\right)$ and $\cos^2\left(Kz+\phi\right)$ could lead to extinction.

In the following we will investigate the experimentally relevant $N$-particle case. In the low-energy regime the scattering of a soliton off a scattering potential has been predicted to lead to the creation of a nonlocal quantum superposition of the form (\ref{eq:NOON}). Again the superposition of wave-functions (\ref{eq:solitonwavefunction}) with opposing $K$ leads to an interference pattern
\begin{equation}
\langle \delta \left(x -\frac{x_1 + \dots + x_N}{N}  \right)  \rangle_{\rm{N,CoM}} \propto \cos^2\left( Kx \right)
\end{equation}
with contrast $C_{N,\rm{CoM}}=1$ in the CoM density distribution while the contrast of the single-particle density is given by (\ref{eq:AppA})
\begin{equation}\label{eq:contrastNoneparticle}
C_{N,\rm{one}} = \frac{\pi\sqrt{N-1}\sqrt{\frac{E_{\mathrm{kin}}}{|\mu|}}}{\sinh\left( \pi\sqrt{N-1}\sqrt{\frac{E_{\mathrm{kin}}}{|\mu|}} \right)}.
\end{equation}
As shown in figure~\ref{fig:contrastN}~(b) the contrast only takes on values $C_{N,\rm{one}}\approx 1$ for extremly small ratios $E_{\mathrm{kin}}/|\mu|$ and vanishes otherwise. A comparison of CoM and single-particle density interference patterns is shown for experimentally realistic parameters \cite{WeissCastin09} in figure \ref{fig:contrastN} (c). Though the calculations have been performed for plane waves of the CoM rather than wave-packets it can be deduced that also with wavepackets no interference in the single-particle density could be observed.

\begin{figure}
\includegraphics[width=1.0\columnwidth]{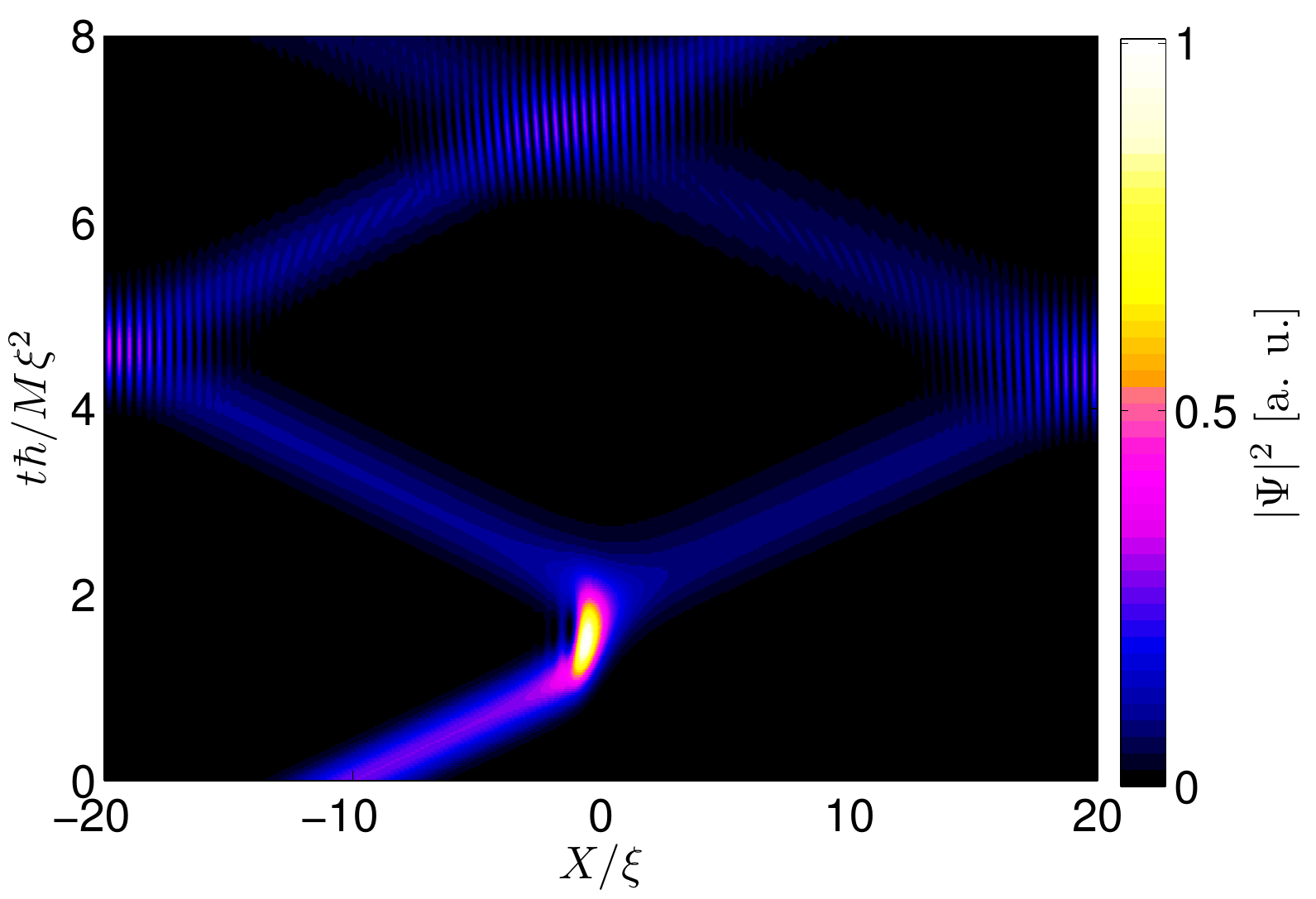}
\caption{\label{fig:timeevolN} Time evolution for the scattering of a soliton consisting of $N=100$ particles at a delta potential in the effective potential approach: numerically, by discretization of CoM Schr\"odinger equation, calculated CoM density distribution versus dimensionless CoM coordinate $X/\xi$ and dimensionless time $t\hbar/\xi^2M$. The initial wave packet with $\Psi(X,t=0)\propto \exp(\mathrm{i}K_0X)\exp(-X^2/a^2)$, where $a/\xi=10/3$, is launched with an initial momentum $K_0\xi=8$ and is centered around $X/\xi=-10$. At about $t\hbar/\xi^2M=1.5$ the wave packet reaches the scattering potential which leads to the creation of a Schr\"odinger-cat state with 50\% - 50 \% splitting. At $t\hbar/\xi^2M=4$ the scattering potential is switched off and the coherent parts recombine and interfere at $t\hbar/\xi^2M\approx7$. Number of lattice points: 1601. The additional reflection potentials from figures~\ref{fig:comdensity} and~\ref{fig:onedensity} are here implemented via reflection on the edges of the lattice.}
\end{figure}

Within the effective potential approach \cite{WeissCastin09} it is possible to numerically investigate the time evolution of the CoM density for the scattering of a soliton consisting of $N=100$ atoms at a delta potential which is shown in figure \ref{fig:timeevolN}. The effective potential~(\ref{eq:Veff}) being the convolution of the scattering potential with the mean-field soliton density~(\ref{eq:rhosoliton}) is again of the form $V_{\mathrm{eff}}\sim 1/\cosh^2\left(X/2\xi\right)$ where $\xi = \hbar^2|g_{1\rm{d}}|N$ is the soliton size. Clearly the splitting of the wave packet due to scattering at the barrier potential can be observed. After switching off the barrier potential the coherent parts of the wave packet recombine and display an interference pattern with high contrast.

Summarising these results we have shown that the interference pattern of the CoM density distribution for the ground state soliton always has maximum contrast both for the dimer and the $N$-particle BMWS. For the dimer the contrast in the single-particle density interference pattern is, depending on the ratio of binding to kinetic energy, reduced but still visible. As shown in figure~\ref{fig:contrastN}~(b) for $N$ particles the interference vanishes for all but experimentally unrealistic values of $E_{\mathrm{kin}}/|\mu|$: for low kinetic energies the transit time $t_{\rm{trans}}$ of the nonlocal superposition states in the interferometer would be too long in comparison to the time-scales of the main source of decoherence: particle losses. For the parameters of \cite{WeissCastin09} where the scattering of a soliton consisting of $N=100$ particles is investigated a typical value would be $t_{\rm{trans}}\lesssim 200$ ms. While one-body losses due to collisions with the background gas could be suppressed by a very high vacuum \cite{AndersonEtAl95} three body-losses can be controlled by the chosen experimental parameters, leading in this case to a negligible loss event probability of 3\%. Additionally, the very low temperatures required for the proposed protocol are experimentally accessible \cite{LeanhardtEtAl03}.   Besides it is important to note that the number of particles constituting the soliton is experimentally well controllable: a number postselection of $N =100\pm5$ atoms is feasible \cite{privcomGross}. So far, we have assumed the CoM of the soliton to be known with arbitrary precision. Nonetheless, as the CoM of the soliton can be determined an order of magnitude better than the soliton width \cite{privcomOberthaler} the proposed measurement of the CoM density distribution should be experimentally feasible. Not singling out the CoM but averaging over the whole measured density distributions would smear out this interference pattern as can be seen in figure~\ref{fig:contrastN}~(b).

\section{Conclusion}
To summarise, for sufficiently low kinetic energies the scattering of tightly bound dimers or BMWS can lead to the creation of nonlocal mesoscopic quantum superposition states~(\ref{eq:NOON}). For the scattering of a dimer off a delta potential the effective potential approach from \cite{WeissCastin09} could be tested numerically. The approach has been found to describe qualitatively correct the dynamics though the consideration of higher order terms in the perturbation theory leads to clearly improved results. While so far only first order corrections to the effective potential have been investigated \cite{WeissCastin09}, here we have presented a case where higher-order corrections become necessary.

Furthermore, for the two-particle case we have numerically shown that the subsequent interference of the two coherent parts of the wave-function leads to a perfect contrast in the CoM coordinate while the visibility of the interference pattern in the single-particle density is reduced. Our analytical calculations confirm this behaviour. For a tightly bound dimer the contrast will only be reduced whereas for experimentally realistic atom numbers on the order of $N=100$ the contrast in the single-particle density vanishes in all but experimentally unrealistic regimes. Nevertheless, in the CoM density an interference pattern still is clearly displayed.

Concluding, we have shown the occurrence of an interference pattern with high contrast both for the dimer and a soliton consisting of about 100 atoms in the CoM density distribution. Additionally, after the scattering all the particles should be clustered in a single lump, in repeated measurements randomly distributed on either side of the scattering potential. Combining these two characteristics can serve as a signature of nonlocal quantum superposition states.
\paragraph*{Note added in proof:} The effective potential method~\cite{WeissCastin09} was also
independently developed in~\cite{SachaEtAl09}.

\acknowledgments 
We acknowledge discussions with S. Arlinghaus, Y. Castin, S. A. Gardiner, T. Gasenzer, M. Holthaus, R. Hulet and M. Oberthaler. BG thanks C. S. Adams and S. A. Gardiner for hospitality at the University of Durham and acknowledges funding by the 'Studienstiftung des deutschen Volkes' and the 'Heinz Neum\"uller Stiftung'. CW acknowledges funding from UK EPSRC (Grant No. EP/G056781/1).

\begin{appendix}

\section{Effective potential approach for the scattering of a dimer at a delta potential} \label{app:effpot}
The energy change (\ref{eq:firstorder}) in first order perturbation theory is
\begin{eqnarray}\label{eq:help1}
\Delta E^{(1)}(X) &= \int_0^{\infty} \mathrm{d}x|\Psi^{(0)}_{0,N=2}|^2V_X\nonumber \\
&= -2v_0 \delta(X) + 4\beta v_0 \exp(-4\beta |X|),
\end{eqnarray}
leading to the effective potential~(\ref{eq:effpot1}). For compactness we omit the argument in $V_X(X)$.\\
For higher order corrections the excited states~(\ref{eq:CoMandrelcoord}) for a dimer are required. Their relative part reads
\begin{equation} \label{eq:rel}
\Psi_{k,\mathrm{rel}}^{(0)}(x_{\rm{r}})= 2\sqrt{\mathcal{N}_k}\left[\cos\left(kx_{\rm{r}} \right)-\frac{\beta}{k} \sin\left(k|x_{\rm{r}}|\right) \right].
\end{equation}
Here the particles have been enclosed in a fictitious box of size $L$, later assumed to be infinitely large. Using periodic boundary conditions this leads to the norm
\begin{equation}
\mathcal{N}_k = \frac{1}{2L\left(1+\frac{\beta^2}{k^2} \right)}.
\end{equation}
In Rayleigh-Schr\"odinger perturbation theory with continous eigenstates~(\ref{eq:rel}) the energy correction in second order is given by
\begin{equation}
\Delta E^{(2)}(X) = \int_0^{\infty} \mathrm{d}k g(k) \frac{\left|\int_0^L \mathrm{d}x {\Psi_{0,N=2}^{{(0)^{\ast}}}}V_X\Psi_{k,\mathrm{rel}}^{(0)}\right|^2}{E_{0}^{(0)}(2)-E^{(0)}_k(2)} 
\end{equation}
where $g(k) = \frac{L}{2\pi}$ and the argument $x_{\rm{r}}$ in (\ref{eq:rel}) is again omitted. The integral is restricted to the interval $0<k<\infty$ as the relative part of the wave-function (\ref{eq:rel}) is symmetric with respect to $k$. With $E_0^{(0)}(N=2) = -\frac{\hbar^2\beta^2}{m}$, $E_k^{(0)}(N=2) = \frac{\hbar^2k^2}{m}$,
\begin{align} \label{eq:help2}
\left|\int_0^L \mathrm{d}x {\Psi_{0,N=2}^{{(0)^{\ast}}}}V_X\Psi_{k,\mathrm{rel}}^{(0)}\right|^2 =& 32\beta v_0^2 \mathcal{N}\exp(-4\beta |X|) \nonumber \\ &\times\left[\cos(2k|X|) - \frac{\beta}{k}\sin(2k|X|)  \right]^2
\end{align}
and using the Gradsteyn integrals (3.264), (3.773) and (3.264) we get
\begin{equation}
\Delta E^{(2)}(X) = -\frac{2v_0^2 m}{\hbar^2}\exp(-4\beta|X|)\left(1-8\beta|X|\exp(-4\beta|X|) \right).
\end{equation}
corresponding to the contribution $V_{\rm{eff}}^{(2)}$ to the effective potential in second order, given by equation (\ref{eq:Veff2}). 
The energy correction in third order is given by
\begin{widetext}
\begin{eqnarray}
E_0^{(3)}(X)=&\frac{L^2}{4\pi^2}\int\limits_0^{\infty} \mathrm{d}k_1\int\limits_0^{\infty}\mathrm{d}k_2 \frac{\langle \Psi_{0,N=2}^{(0)}|V_X|\Psi_{k_1,\rm{rel}}^{(0)}\rangle \langle \Psi_{k_1,\rm{rel}}^{(0)}|V_X|\Psi_{k_2,\rm{rel}}^{(0)}\rangle  \langle \Psi_{k_2,\rm{rel}}^{(0)}|V_X|\Psi_{0,N=2}^{(0)}\rangle}{\left[E_0(2) - E_{k_1}(2) \right]\left[E_0(2) - E_{k_2}(2) \right]} \nonumber \\ & -\frac{L}{2\pi} \int_0^{\infty} \mathrm{d}k_1 \frac{|\langle \Psi_{0,N=2}^{(0)}|V_X|\Psi_{k_1,\rm{rel}}^{(0)}\rangle |^2\langle \Psi_{0,N=2}^{(0)}|V_X|\Psi_{0,\rm{rel}}^{(0)}\rangle }{\left[ E_0(2) - E_{k_1}(2) \right]^2}\nonumber 
\end{eqnarray}
\end{widetext}
With equations (\ref{eq:help1}), (\ref{eq:help2}) and
\begin{align}
\langle \Psi_{k_1,\rm{rel}}^{(0)}|V_X|\Psi_{k_2,\rm{rel}}^{(0)}\rangle  = & \frac{4v_0}{L\left(1+\frac{\beta^2}{k_1^2}\right)^{(1/2)} \left(1+\frac{\beta^2}{k_2^2}\right)^{(1/2)}} \nonumber \\&\times\left[\cos(2k_1|X|) - \frac{\beta}{k_1}\sin(2k_1 |X|)  \right] \nonumber \\
& \times\left[ \cos(2k_2|X|) - \frac{\beta}{k_2}\sin(2k_2 |X|) \right]\hspace{0.4cm}
\end{align}
as well as the Gradsteyn integrals (3.773) and (3.251) this gives
\begin{align}
E_0^{(3)}(X) = & 16v_0^3 \frac{m^2}{\hbar^4}\exp(-4\beta|X|) \nonumber \\& \times\left[\frac{1}{16\beta} - \left(|X|+\frac{1}{4\beta} \right)\exp(-4\beta|X|)\right.  \nonumber  \\ &+\left(8|X|^2\beta +|X| +\frac{1}{8\beta}\right)\exp(-8\beta|X|) \nonumber \\ &\left.+\left(\frac{1}{8\beta^2} - \frac{1}{16\beta^2}\exp(-4\beta|X|)\right)\delta\left(X\right) \right]. 
\end{align}
This corresponds to the third order correction $V_{\rm{eff}}^{(3)}$ to the effective potential, cf. (\ref{eq:Veff3}).

\section{Interference patterns}\label{eq:AppA}
\subsection{Calculation of the interference pattern for the excited states (\ref{eq:excstatespm}) in the two-particle case}
For the wave function $\Psi_+$ with $K$ and $k$ the corresponding wave function $\Psi_-$ has been obtained by the replacements $K\rightarrow -K$ and $k \rightarrow -k$. With the superposition state 
\begin{equation}
|\Psi_+\rangle + |\Psi_-\rangle = 4\sqrt{\mathcal{N}_k}\left[\cos\left(kx_{\rm{r}} \right)-\frac{\beta}{k} \sin\left(k|x_{\rm{r}}|\right) \right]\cos\left(KX\right)
\end{equation}
this leads to a CoM density distribution
\begin{align}
\lim_{L \to \infty}\langle \delta\left(x-X \right) \rangle_{\rm{exc}} &=  \lim_{L \to \infty}\int_0^{L}\mathrm{d}x_{\rm{r}}\int_{-L/2}^{L/2}\mathrm{d}X 16\mathcal{N}_k \nonumber \\ &\;\;\;\;\; \times\left[\cos\left(k|x_{\rm{r}}| \right)-\frac{\beta}{k} \sin\left(k|x_{\rm{r}}|\right) \right]^2\cos^2\left(KX\right)\nonumber \\ &=4 \cos^2\left(Kx\right).
\end{align}

\subsection{$N$-particle case}
The interference pattern in the CoM density under consideration of equation~(\ref{eq:rhosoliton}) is
\begin{eqnarray}
& \langle \delta \left(x -\frac{x_1 + \dots + x_N}{N}  \right)  \rangle_{N,\rm{CoM}} \nonumber \\
 & =  4\int \mathrm{d}x_1\cdots\mathrm{d}x_N |\Psi_{0,\mathrm{rel}}|^2\cos^2\left(K \frac{x_1 + \dots + x_N}{N}\right)  \nonumber \\&\;\;\;\;\;\;\;\times \delta \left(x -\frac{x_1 + \dots + x_N}{N} \right) \nonumber \\
& =  4 \int_{-\infty}^{\infty} \mathrm{d}y \cos^2\left(Kx\right)\int \mathrm{d}x_1\cdots \mathrm{d}x_N |\Psi_{0,\mathrm{rel}}|^2  \nonumber \\&\;\;\;\;\;\;\;\times\delta \left(x -\frac{x_1 + \dots + x_N}{N} \right)\delta \left(x_1-y \right) \nonumber \\
& =  \frac{mg_{1\rm{d}}}{\hbar^2}N^2\cos^2\left(Kx\right)\int_{-\infty}^{\infty} \mathrm{d}y \left[\cosh\left( \frac{mg_{1\rm{d}}}{2\hbar^2}N\left(y-x  \right)  \right)  \right]^{-2} \nonumber \\
& =  4N\cos^2\left( Kx \right).
\end{eqnarray}
The interference pattern in the single-particle density (\ref{eq:onedensity}) is, again using equation~(\ref{eq:rhosoliton}) and Gradsteyn (3.982), given by
\begin{eqnarray}
& \langle \delta\left( x - x_1  \right)  \rangle_{N,\rm{one}} \nonumber \\ & =  \int_{-\infty}^{\infty} \mathrm{d}X \cos^2\left(KX \right)\int \mathrm{d}x_1\cdots \mathrm{d}x_N|\Psi_{0,\mathrm{rel}}|^2 \nonumber \\&\;\;\;\;\;\;\;\times \delta \left(X-\frac{x_1 + \dots + x_N}{N}  \right)\delta\left( x - x_1  \right) \nonumber \\
& =  2\beta N^2 \int_{-\infty}^{\infty}\mathrm{d}X\frac{\cos^2\left(KX\right)}{\cosh^2\left( \beta N \left(X-x\right) \right)} \nonumber \\
& =  \beta N^2 \int_{-\infty}^{\infty}\mathrm{d}\tilde{X}\frac{1+\cos\left(2K\left(\tilde{X} +x \right)  \right)}{\cosh^2\left( \beta N \tilde{X} \right)} \nonumber \\
& =  2N + \frac{2K\pi}{\beta}\frac{1}{\sinh\left( \frac{K\pi}{\beta N} \right)}\cos\left( 2Kx \right)
\end{eqnarray}

Thus, the intensity can take on the following maximal and minimal values
\begin{eqnarray}
I_{\mathrm{max}} & = \langle \delta\left( 0 - x_1 \right) \rangle = 2N + \frac{2K\pi}{\beta}\frac{1}{\sinh\left(\frac{K\pi}{\beta N} \right)}, \\
I_{\mathrm{min}} & =  \langle \delta\left( \frac{\pi}{2K} - x_1 \right) \rangle = 2N - \frac{2K\pi}{\beta}\frac{1}{\sinh\left(\frac{K\pi}{\beta N} \right)}.
\end{eqnarray}
This leads to the contrast
\begin{equation}\label{eq:contrastNone}
C_{N,\rm{one}} = \frac{\frac{K\pi}{\beta N}}{\sinh\left(\frac{K\pi}{\beta N}\right)}.
\end{equation}
Using equation~(\ref{eq:gap}) and (\ref{eq:energysol}) gives expression (\ref{eq:contrastNoneparticle}) for the contrast in dependence of 
\begin{equation}
\frac{E_{\mathrm{kin}}}{|\mu|} = \frac{K^2}{\beta^2 N^2(N-1)}.
\end{equation}
\end{appendix}


\end{document}